\documentclass[english]{llncs}
\usepackage[T1]{fontenc}
\usepackage[latin1]{inputenc}
\usepackage{subfigure}
\usepackage{graphicx}
\usepackage{setspace}

\makeatletter

\providecommand{\tabularnewline}{\\}

\def\draftversion{N}                
\def\note[#1]#2{\message{(#1)}\if\draftversion
{\noindent\em[#2]\/}\fi}
\if \draftversion Y
\fi

\usepackage{babel}
\makeatother
\begin{document}

\vfill{}
\title{Fast Tuning of Intra-Cluster Collective Communications}
\vfill{}

\author{Luiz Angelo Barchet-Estefanel\thanks{Supported by grant BEX 1364/00-6 from CAPES - Brazil},
Grégory Mounié}

\institute{Laboratoire ID - IMAG, Project APACHE\thanks{This project is supported by CNRS, INPG, INRIA and UJF}
\\
51, Avenue Jean Kuntzmann, F38330 Montbonnot St. Martin, France\\
 \email{\{Luiz-Angelo.Estefanel,Gregory.Mounie\}@imag.fr}}

\maketitle
\begin{abstract}
Recent works try to optimise collective communication in grid systems
focusing mostly on the optimisation of communications among different
clusters. We believe that intra-cluster collective communications
should also be optimised, as a way to improve the overall efficiency
and to allow the construction of multi-level collective operations.
Indeed, inside homogeneous clusters, a simple optimisation approach
rely on the comparison from different implementation strategies, through
their communication models. In this paper we evaluate this approach,
comparing different implementation strategies with their predicted
performances. As a result, we are able to choose the communication
strategy that better adapts to each network environment.
\end{abstract}

\section{\label{sec:Introduction}Introduction}

The optimisation of collective communications in grids is a complex
task because the inherent heterogeneity of the network forbids the
use of general solutions. Indeed, the optimisation cost can be fairly
reduced if we consider grids as interconnected islands of homogeneous
clusters, if we can identify the network topology.

Most systems only separate inter and intra-cluster communications,
optimising communication across wide-area networks, which are usually
slower than communication inside LANs. Some examples of this {}``two-layered''
approach include ECO \cite{key-5}, MagPIe \cite{key-6,key-9} and
even LAM-MPI 7 \cite{key-32}. While ECO and MagPIe apply this concept
for wide-area networks, LAM-MPI 7 applies it to SMP clusters, where
each SMP machine is an island of fast communication. Even though,
there is no real restriction on the number of layer and, indeed, the
performance of collective communications can still be improved by
the use of multi-level communication layers, as observed by \cite{key-31}. 

If most works today use the {}``islands of clusters'' approach,
to our knowledge none of them tries to optimise the intra-cluster
communication. We believe that while inter-cluster communication represents
the most important aspect in grid-like environments, intra-cluster
optimisation also should be considered, specially if the clusters
should be structured in multiple layers \cite{key-31}. In fact, collective
communications in local-area networks can still be improved with the
use of message segmentation \cite{key-33,key-9} or the use of different
communication strategies \cite{key-34}.

In this paper we propose the use of well known techniques for collective
communication, that due to the relative homogeneity inside each cluster,
may reduce the optimisation cost. Contrarily to \cite{key-10}, we
decided to model the performance of different implementation strategies
for collective communications and to select, according to the network
characteristics, the most adapted implementation technique for each
set of parameters (communication pattern, message size, number of
processes). Hence, in this paper we illustrate our approach with two
examples, the Broadcast and Scatter operations, and we validate our
approach by comparing the performance from real communications and
the models' predictions.

The rest of this paper is organised as follows: Section \ref{sec:System_Model}
presents the definitions and the test environment we will consider
along this paper. Section \ref{sec:pLogP models} presents the communication
models we developed for both Broadcast's and Scatter's most usual
implementations. In Section \ref{sec:Practical-Results} we compare
the predictions from the models with experimental results. Finally,
Section \ref{sec:Conclusions} presents our conclusions, as well as
the future directions of the research.

\section{\label{sec:System_Model}System Model and Definitions}

In this paper we model collective communications using the \emph{parameterised
LogP} model, or simply pLogP \cite{key-9}. Hence, all along this
paper we shall use the same terminology from pLogP's definition, such
as \emph{g(m)} for the gap of a message of size \emph{m}, \emph{L}
as the communication latency between two nodes, and \emph{P} as the
number of nodes. In the case of message segmentation, the segment
size \emph{s} of the message \emph{m} is a multiple of the size of
the basic datatype to be transmitted, and it splits the initial message
\emph{m} into \emph{k} segments. Thus, \emph{g(s)} represents the
gap of a segment with size \emph{s}.

The pLogP parameters used to feed our models were previously obtained
with the MPI LogP Benchmark tool \cite{key-7} using LAM-MPI 6.5.9
\cite{key-15}. The experiments to obtain pLogP parameters, as well
as the practical experiments, were conducted on the \textbf{ID/HP
icluster-1} from the ID laboratory Cluster Computing Centre%
\footnote{http://www-id.imag.fr/Grappes/%
}, with 50 Pentium III machines (850Mhz, 256MB) interconnected by a
switched Ethernet 100 Mbps network.

\section{\label{sec:pLogP models}Communication Models with pLogP}

Due to the limited space, we cannot present models for all collective
communication, thus we chose to present the Broadcast and the Scatter
operations. Although they are two of the simplest collective communication
patterns, practical implementations of MPI usually construct other
collective operations, as for example, Barrier, Reduce and Gather,
in a very similar way, what makes these two operations a good example
for our models accuracy. Further, the optimisation of grid-aware collective
communications explores intensively such communication patterns, as
for example the AllGather operation in MagPIe, which has three steps:
a Gather operation inside each cluster, an AllGatherv among the clusters'
roots and a Broadcast to the cluster's members.

\subsection{\label{sub:Broadcast}Broadcast}

With Broadcast, a single process, called \emph{root,} sends the same
message of size \emph{m} to all other $(P-1)$ processes. Among the
classical implementations for broadcast in homogeneous environments
we can find flat, binary and binomial trees, as well as chains (or
pipelines). It is usual to apply different strategies within these
techniques according to the message size, as for example, the use
of a \emph{rendezvous} message that prepares the receiver to the incoming
of a large message, or the use of non-blocking primitives to improve
communication overlap. Based on the models proposed by \cite{key-9},
we developed the communication models for some current techniques
and their {}``flavours'', which are presented on Table \ref{table:bcast_models}. 

We also considered message segmentation \cite{key-9,key-34}, which
may improve the communication performance under some specific situations.
An important aspect, when dealing with message segmentation, is to
determine the optimal segment size. Too little messages pay more for
their headers than for their content, while too large messages do
not explore enough the network bandwidth. Hence, we can use the communication
models presented on Table \ref{table:bcast_models} to search the
segment size \emph{s} that minimises the communication time in a given
network. Once determined this segment size \emph{s}, large messages
can be split into $\lfloor m/s\rfloor$ segments, while smaller messages
will be transmitted without segmentation.

As most of these variations are clearly expensive, we did not consider
them on the experiments from Section \ref{sec:Practical-Results},
and focused only in the comparison of the most efficient techniques,
the Binomial and the Segmented Chain Broadcasts.

\begin{table}

\caption{\label{table:bcast_models}Communication Models for Broadcast}

\begin{onehalfspace}
\begin{center}\begin{tabular}{|c|c|}
\hline 
\textbf{Implementation Technique}&
\textbf{Communication Model}\tabularnewline
\hline
\hline 
Flat Tree&
$(P-1)\times g(m)+L$\tabularnewline
\hline 
Flat Tree Rendezvous&
$(P-1)\times g(m)+2\times g(1)+3\ \times L$\tabularnewline
\hline 
Segmented Flat Tree&
$(P-1)\times(g(s)\times k)+L$\tabularnewline
\hline 
Chain&
$(P-1)\times(g(m)+L)$\tabularnewline
\hline 
Chain Rendezvous&
$(P-1)\times(g(m)+2\times g(1)+3\ \times L)$\tabularnewline
\hline 
Segmented Chain (Pipeline)&
$(P-1)\times(g(s)+L)+(g(s)\times(k-1))$\tabularnewline
\hline 
Binary Tree&
$\leq\lceil log_{2}P\rceil\times(2\times g(m)+L)$\tabularnewline
\hline 
Binomial Tree&
$\lfloor log_{2}P\rfloor\times g(m)+\lceil log_{2}P\rceil\times L$\tabularnewline
\hline 
Binomial Tree Rendezvous&
$\lfloor log_{2}P\rfloor\times g(m)+\lceil log_{2}P\rceil\times(2\times g(1)+3\times L)$\tabularnewline
\hline 
Segmented Binomial Tree&
$\lfloor log_{2}P\rfloor\times g(s)\times k+\lceil log_{2}P\rceil\times L$\tabularnewline
\hline
\end{tabular}\end{center}
\end{onehalfspace}

\vspace{-0.8cm}
\end{table}

\subsection{\label{sub:Scatter}Scatter}

The Scatter operation, which is also called {}``personalised broadcast'',
is an operation where the \emph{root} holds \emph{$m\times P$} data
items that should be equally distributed among the P processes, including
itself. It is believed that optimal algorithms for homogeneous networks
use flat trees \cite{key-9}, and by this reason, the Flat Tree approach
is the \emph{default} Scatter implementation in most MPI implementations.
The idea behind a Flat Tree Scatter is that, as each node shall receive
a different message, the root shall sends these messages directly
to each destination node. 

To better explore our approach, we constructed the communication model
for other strategies (Table \ref{table:Models-for-Scatter}) and,
in this paper, we compare Flat Scatter and Binomial Scatter in real
experiments. In a first look, a Binomial Scatter is not as efficient
as the Flat Scatter, because each node receives from the parent node
its message as well as the set of messages it shall send to its successors.
On the other hand, the cost to send these {}``combined'' messages
(where most part is useless to the receiver and should be forwarded
again) may be compensate by the possibility to execute parallel transmissions.
As the trade-off between transmission cost and parallel sends is represented
in our models, we can evaluate the advantages of each model according
to the clusters' characteristics. 

\begin{table}

\caption{\label{table:Models-for-Scatter}Communication Models for Scatter}

\begin{center}\begin{tabular}{|c|c|}
\hline 
\textbf{Implementation Technique}&
\textbf{Communication Model}\tabularnewline
\hline
\hline 
Flat Tree&
$(P-1)\times g(m)+L$\tabularnewline
\hline 
Chain&
$\sum_{j=1}^{P-1}g(j\times m)+(P-1)\times L$\tabularnewline
\hline 
Binomial Tree&
$\sum_{j=0}^{\lceil log_{2}P\rceil-1}g(2^{j}\times m)+\lceil log_{2}P\rceil\times L$\tabularnewline
\hline
\end{tabular}\end{center}
\end{table}

\vspace{-1cm}

\section{\label{sec:Practical-Results}Practical Results}

\subsection{\label{sub:Broadcast_practical}Broadcast}

\begin{figure}[h]
\begin{center}\subfigure[Binomial Tree]{\includegraphics[%
  bb=0bp 20bp 378bp 221bp,
  width=0.52\linewidth,
  keepaspectratio]{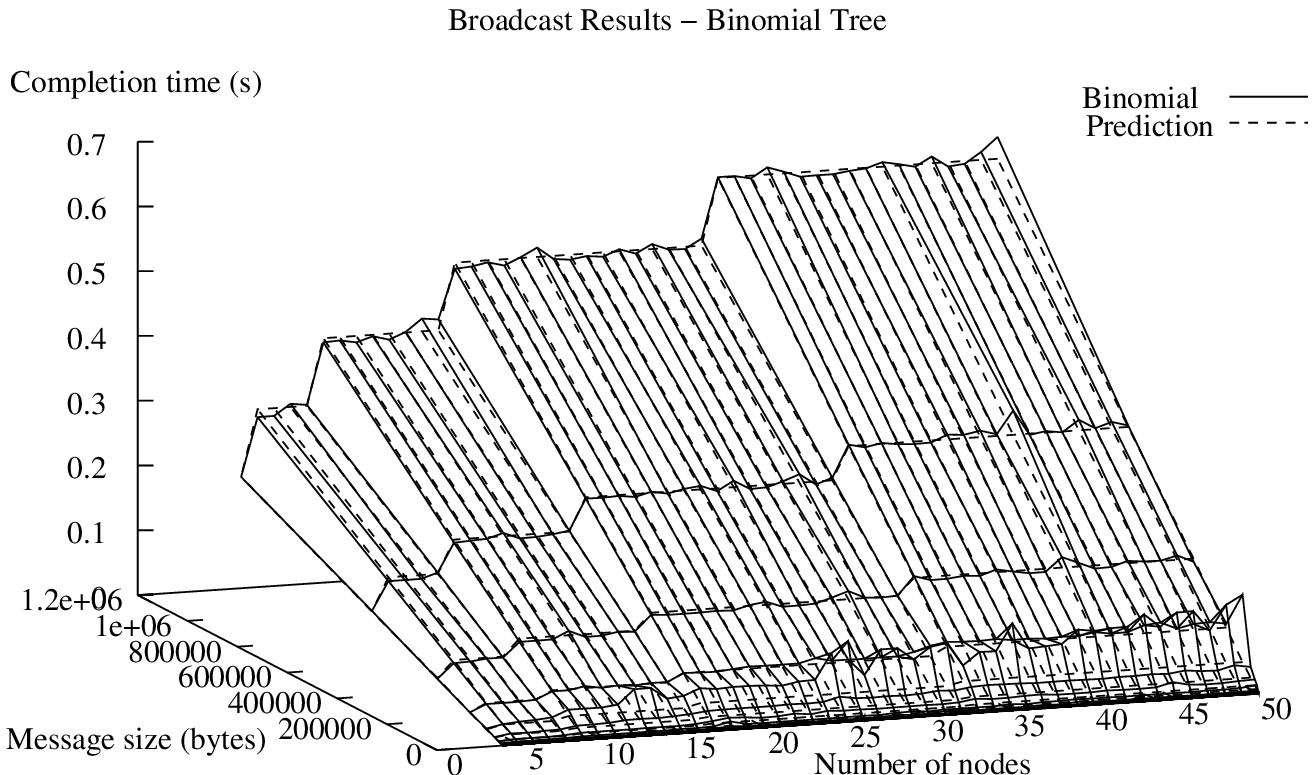}}\subfigure[Segmented Chain - 8kB segments]{\includegraphics[%
  bb=0bp 20bp 388bp 226bp,
  width=0.52\linewidth]{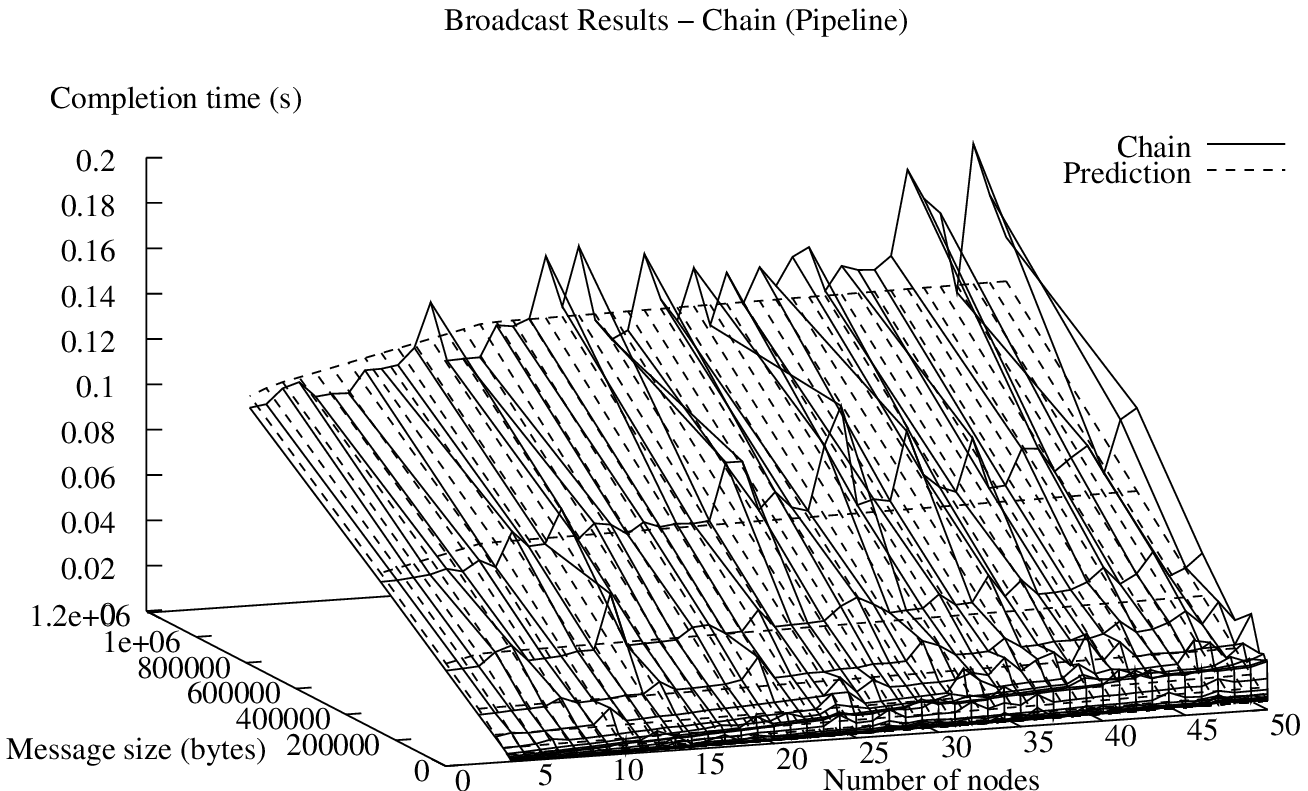}}\end{center}

\vspace{-0.5cm}

\caption{\label{Figure:Comparison-Bcasts}Comparison between models and real
results}
\end{figure}

To evaluate the accuracy of our optimisation approach, we measured
the completion time of the Binomial and the Segmented Chain Broadcasts,
and we compared these results with the model predictions. Through
the analysis of Figs. \ref{Figure:Comparison-Bcasts}(a) and \ref{Figure:Comparison-Bcasts}(b),
we can verify that models' predictions follow closely the real experiments.
Indeed, both experiments and models predictions show that the Segmented
Chain Broadcast is the most adapted strategy to our network parameter,
and consequently, we can rely on the models' predictions to chose
the strategy we will apply. 

Although models were accurate enough to select the best adapted strategy,
a close look at the Fig. \ref{Figure:Comparison-Bcasts} still shows
some differences between model's predictions and the real results.
We can observe that, in the case of the Binomial Broadcast, there
is a non expected delay when messages are small. In the case of the
Segmented Chain Broadcast, however, the execution time is slightly
larger than expected. Actually, we believe that both variations derive
from the same problem.

\begin{figure}[h]
\begin{center}\includegraphics[%
  bb=0bp 20bp 467bp 275bp,
  width=0.45\linewidth,
  keepaspectratio]{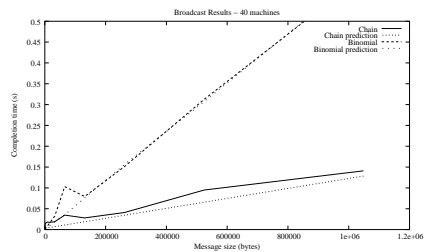}\end{center}

\vspace{-0.5cm}

\caption{\label{Figure:Comparison-between-models Bcast}Comparison between
Chain and Binomial Broadcast}
\end{figure}

Hence, we present in Fig. \ref{Figure:Comparison-between-models Bcast}
the comparison of both strategies and their predictions for a fixed
number of machines. We can observe that predictions for the Binomial
Broadcast fit with enough accuracy the experimental results, except
in the case of small messages (less than 128kB). Actually, similar
discrepancies were already observed by the LAM-MPI team, and according
to \cite{key-14,key-4}, they are due to the TCP acknowledgement policy
on Linux that may delay the transmission of some small messages even
when the TCP\_NODELAY socket option is active (actually, only one
every \emph{n} messages is delayed, with \emph{n} varying from kernel
to kernel implementation).

In the case of the Segmented Chain Broadcast, however, this phenomenon
affects all message sizes. Because large messages are split into small
segments, such segments suffers from the same transmission delays
as the Binomial Broadcast with small messages. Further, due to the
Chain structure, a delay in one node is propagated until the end of
the chain. Nevertheless, the transmission delay for a large message
(and by consequence, a large number of segments) does not increases
proportionally as it would be expected, but remains constant. 

We believe that because these transmission delays are related to the
buffering policy from TCP, we believe that the first segments that
arrive are delayed by the TCP acknowledge policy, but the successive
arrival of the following segments forces the transmission of the remaining
segments without any delay.

\subsection{\label{sub:Scatter_practical}Scatter}

\begin{figure}[h]
\vspace{-0.5cm}

\begin{center}\subfigure[Flat Tree Scatter]{\includegraphics[%
  bb=0bp 20bp 499bp 271bp,
  width=0.52\linewidth]{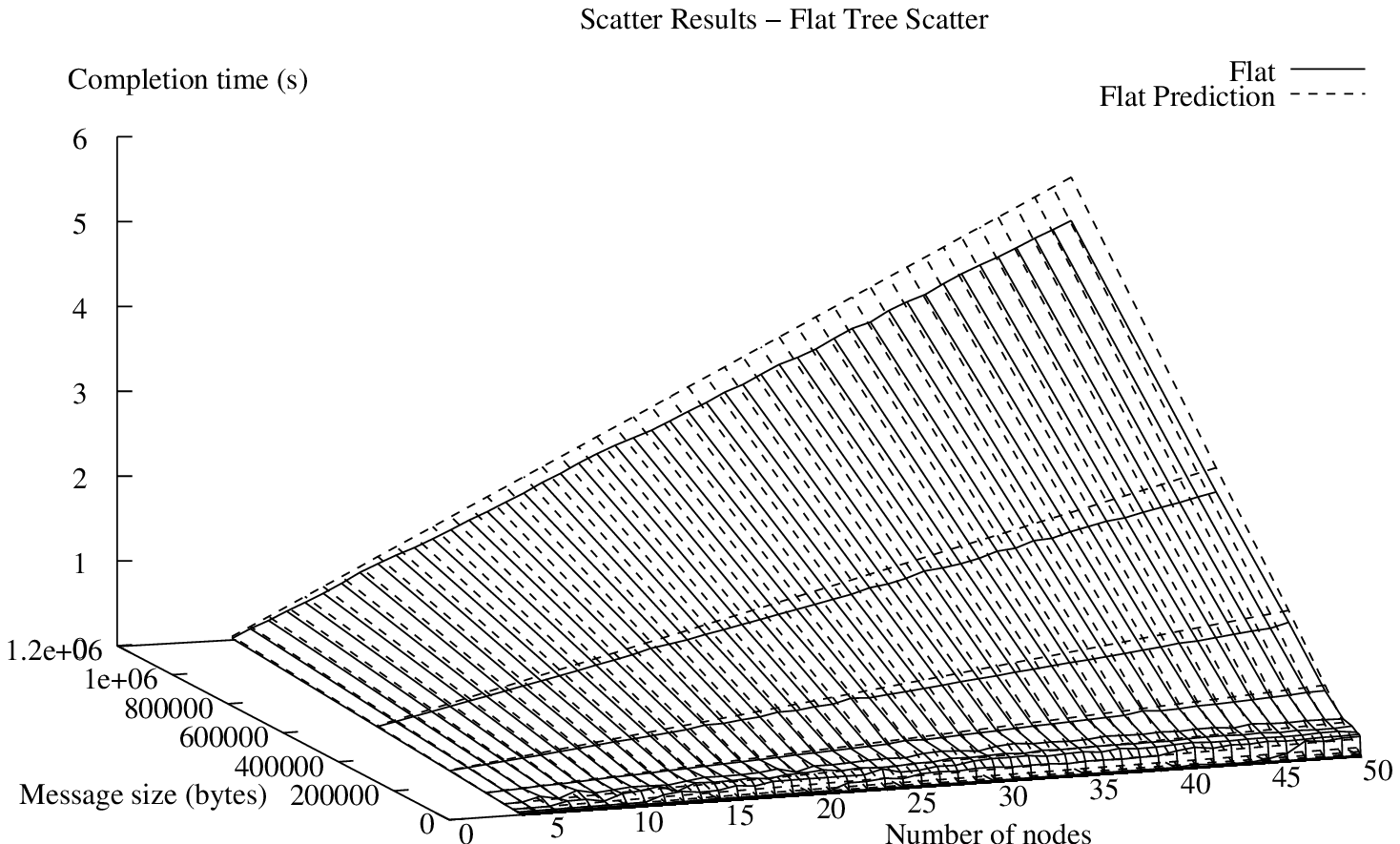}}\subfigure[Binomial Tree Scatter]{\includegraphics[%
  bb=0bp 20bp 499bp 271bp,
  width=0.52\linewidth]{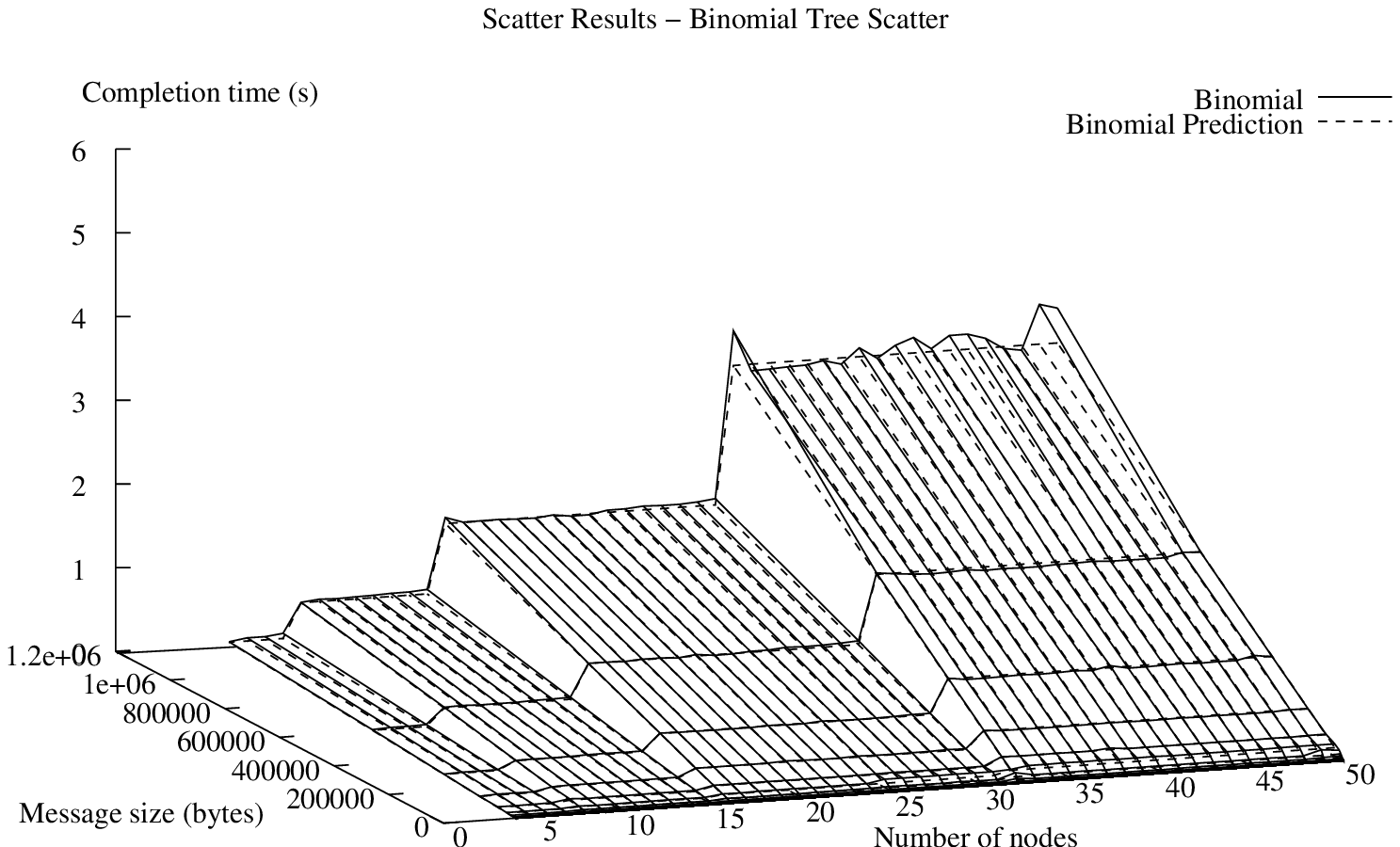}}\vspace{-0.5cm}\end{center}

\caption{\label{Figure:Comparison-Scatter}Comparison between models and real
results}
\end{figure}

In the case of Scatter, we compare the experimental results from Flat
and Binomial Scatters with the predictions from their models. Due
to our network characteristics, our experiments shown that a Binomial
Scatter can be more efficient than Flat Scatter, a fact that is not
usually explored by traditional MPI implementations. As a Binomial
Scatter should balance the cost of combined messages and parallel
sends, it might occur, as in our experiments, that its performance
outweighs the {}``simplicity'' from the Flat Scatter with considerable
gains according to the message size and number of nodes, as shown
in Figs. \ref{Figure:Comparison-Scatter}(a) and \ref{Figure:Comparison-Scatter}(b).
In fact, the Flat Tree model is limited by the time the root needs
to send successive messages to different nodes (the gap), while the
Binomial Tree Scatter depends mostly on the number of nodes, which
defines the number of communication steps through the $\lceil log_{2}P\rceil\times L$
factor. These results show that the communication models we developes
are accurate enough to identify which implementation is the best adapted
to a specific environment and a set of parameters (message size, number
of nodes).

\begin{figure}[h]
\begin{center}\includegraphics[%
  bb=0bp 20bp 467bp 275bp,
  width=0.45\linewidth]{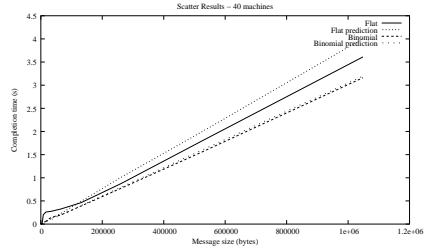}\end{center}

\vspace{-0.5cm}

\caption{\label{Figure:Comparison-between-models-Scatter}Comparison between
Flat and Binomial Scatter}
\end{figure}

Further, although we can observe some delays related to the TCP acknowledgement
policy on Linux when messages are small, specially in the Flat Scatter,
these variations are less important than those from the Broadcast,
as depicted in Fig. \ref{Figure:Comparison-between-models-Scatter}. 

What called our attention, however, was the performance of the Flat
Tree Scatter, that outperformed our predictions, while the Binomial
Scatter follows the predictions from its model. We think that the
multiple transmissions from the Flat Scatter become a ''bulk transmission'',
which forces the communication buffers to transfer the successive
messages all together, somehow similarly to the successive sends on
the Segmented Chain Broadcast. Hence, we observe that the pLogP parameters
measured by the pLogP benchmark tool are not adapted to such situations,
as it considers only individual transmissions, mostly adapted to the
Binomial Scatter model. 

This behaviour seems to indicate a relationship between the number
of successive messages sent by a node and the buffer transmission
delay, which are not considered in the pLogP performance model. As
this seem a very interesting aspect for the design of accurate communication
models, we shall closely investigate and formalise this ''multi-message''
behaviour in a future work.

\section{\label{sec:Conclusions}Conclusions and Future Works}

Existing works that explore the optimisation of heterogeneous networks
usually focus only the optimisation of inter-cluster communication.
We do not agree with this approach, and we suggest to optimise both
inter-cluster and intra-cluster communication. Hence, in this paper
we described how to improve the communication efficiency on homogeneous
cluster through the use of well known implementation strategies. 

To compare different implementation strategies, we rely on the modelling
of communication patterns. Our decision to use communication models
allows a fast and accurate performance prediction for the collective
communication strategies, giving the possibility to choose the technique
that best adapts to each environment. Additionally, because the intra-cluster
communication is based on static techniques, the complexity on the
generation of optimal trees is restricted only to the inter-cluster
communication.

Nonetheless, as our decisions rely on network models, their accuracy
needs to be evaluated. Hence, in this paper we presented two examples
that compare the predicted performances and the real results. We shown
that the selection of the best communication implementation can be
made with the help of the communication models. Even if we found some
small variations in the predicted data for small messages, these variations
were unable to compromise the final decision, and we could identify
the probable origin from these variations. Hence, one of our future
works include a deep investigation on the factors that lead to such
variations, and in special the relationship between the number of
successive messages and the transmission delay, formalising it and
proposing extensions to the pLogP model.

In parallel, we will evaluate the accuracy of our models with other
network interconnections, specially Ethernet 1Gb and Myrinet, and
study how to reflect the presence of multi-processors and multi-networks
(division of traffic) in our models. Our research will also include
the automatic discovery of the network topology and the construction
of optimised inter-cluster trees that work together with efficient
intra-cluster communication.

\end{document}